\documentclass[nofootinbib,twocolumn,aps,pre,preprintnumbers,floatfix,amsmath,amssymb]{revtex4-1} 
\usepackage{graphicx,color,amsmath,amssymb,bm}

\newcommand{\half}{\frac{1}{2}}
\newcommand{\e}{\mathrm{e}}
\newcommand{\D}{\mathrm{d}}
\newcommand{\vecr}{{\bf r}}

\newcommand{\kbt}{k_{\mathrm{B}}T}

\newcommand{\lb}{\ell_\mathrm{B}}
\newcommand{\ld}{\lambda_\mathrm{D}}
\newcommand{\kd}{\kappa_\mathrm{D}}
\newcommand{\lgc}{\ell_\mathrm{GC}}
\newcommand{\lsig}{\ell_\mathrm{\sigma}}

\begin{document}

\author{Tomer Markovich}
\author{David Andelman}
\affiliation{Raymond and Beverly Sackler School of Physics and Astronomy, Tel Aviv University, Ramat Aviv, Tel Aviv 69978, Israel}

\author{Rudi Podgornik}
\affiliation{Department of Theoretical Physics, J. Stefan Institute,
and \\
Department of Physics, Faculty of Mathematics and Physics\\ University of Ljubljana, 1000 Ljubljana, Slovenia}

\title{Surface Tension of Acid Solutions: Fluctuations beyond the Non-linear Poisson-Boltzmann Theory}

\date{Aug 28, 2016}

\begin{abstract}
We extend our previous study of surface tension of
ionic solutions and apply it to the case of acids (and salts) with strong ion-surface interactions.
These ion-surface interactions yield a non-linear boundary condition with an {\it effective} surface charge due to
adsorption of ions from the bulk onto the interface. The calculation is done using the loop-expansion technique,
where the zero-loop (mean field) corresponds of the non-linear Poisson-Boltzmann equation.
The surface tension is obtained analytically to one-loop order, where the mean-field contribution is a modification of the
Poisson-Boltzmann surface tension, and the one-loop contribution gives a generalization of the Onsager-Samaras result.
Our theory fits well a wide range of different acids and salts,
and is in accord with the reverse Hofmeister series for acids.
\end{abstract}

\maketitle


\section{Introduction}

Solubilization of simple salts in aqueous solutions  increases, in general, its surface tension~\cite{Adamson,Pugh}.
The theoretical foundation of this phenomenon goes back almost a century ago to Wagner~\cite{Wagner},
who suggested an explanation based on image charges (due to the water/air dielectric discontinuity).
Onsager and Samaras (OS), in their {\it tour de force} paper, combined this idea with the Debye-H\"uckel (DH)~\cite{Debye1923} theory,
and calculated the dependence of surface tension
on  salt concentration~\cite{onsager_samaras}.
While being overall successful at low salinity conditions, the OS prediction implies the same increment of the surface
tension for all monovalent salts --- a finding that is at odds with many well-explored physical situations~\cite{Kunz_Book}.
Moreover, some simple monovalent acids and bases not only show quantitative discrepancy with the OS result,
but even act contrary to its qualitative features. These acids and bases may reduce the surface tension
even in the low salinity limit where the OS result is supposed to be universally valid.

A vast number of attempts
that go beyond the OS theory have been proposed and incorporate ion-specific
effects~\cite{Dan2011,Kunz_Book}. They are related to a much broader behavior
of solutes in salt solutions observed already in the late 19th century by
Hofmeister and coworkers~\cite{hofmeister}, known nowadays as the {\it Hofmeister series}. This series emerges in numerous
chemical and biological systems~\cite{collins1985,ruckenstein2003a,kunz2010},
including, but not limited to, forces between mica or silica surfaces~\cite{Sivan2009,Sivan2013,pashely},
as well as surface tension of electrolyte solutions~\cite{air_water_2,air_water_3}.

Over the years, different theoretical approaches were devised to incorporate these experimental
findings into a generalized theoretical framework. Specifically, in
order to incorporate ion-specific interactions, the well-known Poisson-Boltzmann (PB) theory was often taken as a point of departure.
Such an approach, pioneered by Ninham and coworkers~\cite{Ninham1997}, was  later extended by Levin and coworkers~\cite{levin2}.
The Boltzmann weight factor was modified by adding in an {\it ad hoc} manner different types of ion-specific
interactions (assumed to be additive), such as dispersion interactions
~\cite{ninham2001, Edwards2004, LoNostro}, image-charge interaction,  Stern exclusion layer,  ionic cavitation
energy and  ionic polarizability~\cite{levin2}.

The above mentioned modification of the Boltzmann weight factor was used to calculate numerically the surface tension of
electrolytes at the water/air interface,
and with the addition of dispersion forces also at the oil/water interface~\cite{levin2011}.
Similarly, the surface tension of acids~\cite{levin2010} was computed by
taking into account the preferential adsorption of hydrogen (in the form of hydronium ions) to the interface.
We note that while these additional interaction terms may represent
real physical mechanisms underlying the specific ion-surface interactions, these terms are, in general, non-additive~\cite{Kunz_Book}.

In our previous works~\cite{EPL,JCP}, we introduced a self-consistent
phenomenological approach that describes specific ion-surface interactions  in the
form of surface coupling terms in the free energy.
Furthermore, on a formal level, we argue that the original OS result is, in fact, fluctuational in nature,
and it is necessary to extend the PB formalism
to account for fluctuations.
This conceptual and formal development allowed us to derive an analytical theory that
reunites the OS result with the ionic specificity of the Hofmeister series.
Our results demonstrate that simple specific ion-surface interactions
can explain the appearance of the Hofmeister series.

Using the one-loop expansion beyond the linearized Poisson-Boltzmann
theory (the DH theory), we have
obtained~\cite{EPL,JCP} the surface tension dependence on salinity in agreement with experiment, and
with the reverse Hofmeister series. Since this theory is valid only for {\it weak} ion-surface interactions, it is not fully compatible with strongly adhering  ions such as acids.
It is exactly this issue that is addressed in the present work, where
we use a more general approach applicable for both weak and
{\it strong} ion-surface interactions.  We calculate analytically the dependence of the surface tension
on the ionic strength by resorting to the one-loop expansion, while taking into account the full non-linear PB theory.
The extension to strong surface potentials allows us to derive the surface tension
of acids and other strongly adhering charged particles.
Our findings compare favorably with experimental results.

The acids we considered are assumed to be strong.
This means that for a simple monovalent acid dissociated in water,
\begin{equation}
{\rm HX} \leftrightarrows {\rm H}^+ + {\rm X^-}\, ,
\end{equation}
the pK of the acid dissociation reaction is smaller than roughly $-1.5$.
In this case, the ${\rm HX}$ acid is always fully dissociated, irrespective of all the other parameters,
and the H$^+$ concentration is the same as the bulk acid concentration, $[{\rm H}^{+}]=n_b$.
On the contrary, for weak acids, the amount of H$^+$ is smaller than
$n_b$ and depends on $n_b$ as well as on the acid pK.
Treating weak acids is rather a simple extension of the strong acid case, addressed in this paper, if
one takes the pK value to be constant throughout the solution~\cite{weak_acids}.

The outline of the paper is as follows.
In the next section, we present our model (Section~II), calculate the mean-field electrostatic potential and the thermodynamic
grand-canonical potential (Section~II.A), followed by the one-loop correction to the grand potential (Section~II.B).
Section~III includes the surface tension results up to one-loop order, and in Section~IV we compare these
analytical expressions with experiments.
Finally, we draw our conclusions in Section~V.
Appendix~A extends our model to include both adhesivity and fixed surface charges,
while in Appendix~B, we compute the surface tension for strong surface potential and negative anion adhesivity.

\section{The Model}

\begin{figure}[h!]
\includegraphics[scale=0.35,draft=false]{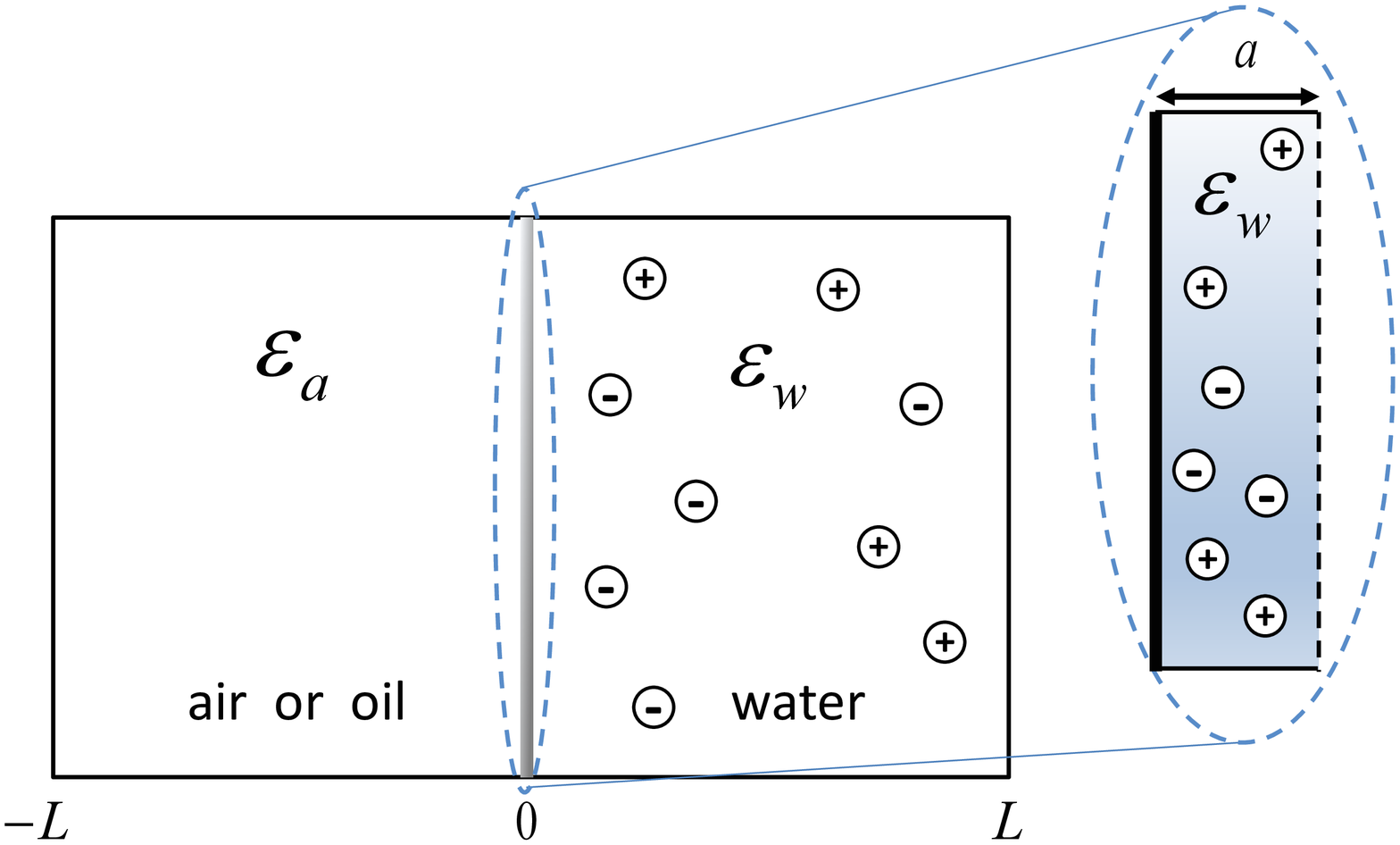}
\caption{\textsf{(color online).
Schematic setup of the system. The aqueous and air phases have the same longitudinal extension, $L$, which
is taken to be macroscopic, $L\rightarrow\infty$.  A small layer proximal to the dividing surface, $0<z<a$, exists
inside the aqueous phase. Within this layer,
the anions and cations interaction with the interface at $z =0$ is modeled by a non-electrostatic potential, $V_{\pm}(z)$. This potential  is zero outside the proximal layer.}}\label{fig1}
\end{figure}

The general problem we consider is the same as in our previous work~\cite{JCP}, composed of aqueous and air phases, as is depicted
schematically in Fig.~\ref{fig1}.
As the full details can be found in Sec.~II of Ref.~\cite{JCP}, only some pertinent highlights of the model are addressed.

We consider a symmetric monovalent ($1$:$1$) electrolyte solution of bulk concentration, $n_b$.
The aqueous phase (water) volume $V=AL$ has a cross-section $A$ and an arbitrary macroscopic length, $L\to \infty$,
with the dividing surface between the aqueous and the air phases at $z=0$.
The two phases are taken as two continuum media with uniform dielectric constants, $\varepsilon_w$ and $\varepsilon_a$, respectively.
We explicitly assume that the ions are confined in the aqueous phase,
due to the large electrostatic self-energy penalty for placing an ion in a low dielectric medium (air or oil).

The model Hamiltonian is:
\begin{eqnarray}
\label{m2}
H = \half\sum_{i,j}q_i q_j u({\bf r}_i,{\bf r}_j)\!-\!\frac{e^2}{2} N u_b\!+\! \sum_{i} V_{\pm}(z_i)\, .
\end{eqnarray}
The first term is the usual Coloumbic interaction, where the summation is done over all the ions in solution,
$q_i=\pm e$ are the charges of monovalent cations and anions,  respectively, and $N$ is the total number of ions in the system.
The second term includes the diverging self-energy, $u_b$, and the last term takes into account
the non-electrostatic ion-surface
potential, $V_{\pm}(z)$. The  potential $V_\pm$ is short ranged and confined to the proximal layer next to the dividing surface, $z \in [0,a]$.
The length $a$ is a microscopic length-scale corresponding to the average ionic size,
or equivalently, to the minimal distance of approach between ions (see Refs.~~\cite{levin2,JCP} for justification).

%

The grand-canonical partition function defined by the above Hamiltonian, Eq.~(\ref{m2}), can be derived in a field theoretical form,
\begin{eqnarray}
\label{e9}
\Xi \equiv \frac{\left(2\pi\right)^{-N/2}}{\sqrt{\det[\beta^{-1} u(\vecr,\vecr')]}}\int {\cal D}\phi\,\e^{-S\left[\phi(\vecr)\right]} \, ,
\end{eqnarray}
where $\beta =1/\kbt$, and $S\left[\phi(\vecr)\right]$ plays the role of a field action,
\begin{eqnarray}
\label{e3a}
\nonumber& S\left[\phi(\vecr)\right] &= \int\D\vecr\,\Bigg( \frac{\beta\varepsilon(\vecr)}{8\pi}[\nabla \phi(\vecr)]^2
- 2\lambda \cos\left[\beta e\phi(\vecr)\right] \Bigg)  \\
\nonumber& &- \lambda\int\D^2r\,\int_{0}^{a}\D z \, \Bigg[\e^{-i\beta e\phi(\vecr)} \left( \e^{- \beta V_{+}(z)} -1 \right)  \\
& & + \, \e^{i\beta e\phi(\vecr)} \left( \e^{- \beta V_{-}(z)} - 1 \right) \Bigg]  \, .
\end{eqnarray}
The derivation of the above equation employs the form of the inverse Coulomb kernel
$u^{-1}(\vecr,\vecr')=-\frac{1}{4\pi} \nabla\cdot[\varepsilon(\vecr)\nabla\delta(\vecr-\vecr')]$,
and the electro-neutrality condition that requires $\lambda_{+} = \lambda_{-} \equiv \lambda$.
The fugacities are defined via the chemical potentials $\mu_\pm$,  where the ion bulk self-energy, $u_b$, is included in their definition,
\begin{eqnarray}
\label{m3a}
\lambda_{\pm} &=& a^{-3}\exp\left(\beta{\mu}_{\pm}\right) \exp\left( \frac{\varepsilon_w}{2} \lb u_b \right) \, ,
\end{eqnarray}
with $\lb = e^2/\varepsilon_w\kbt$ being the Bjerrum length.
The grand potential, $\Omega = -\kbt\ln\Xi$, can be written to first order in a systematic loop expansion, yielding
\begin{eqnarray}
\label{e17}
\nonumber & \beta\Omega & \simeq \beta\Omega_{_{\rm MF}} + \beta\Omega_{_{\rm 1L}} \\
& &= S[\psi] + \half\, {\rm Tr}\ln H_2(\vecr,\vecr') ,
\end{eqnarray}
where the mean-field (MF) term, $\Omega_{_{\rm MF}}$, that depends on the MF electrostatic potential, $\psi(\vecr)$, is derived from the saddle-point equation
\begin{equation}
\frac{\delta S\left[\phi(\vecr)\right]}{\delta \phi(\vecr)}\Big|_{\phi = i \psi} = 0,
\end{equation}
and the Hessian, related to $\Omega_{_{\rm 1L}}$, is defined as
\begin{equation}
H_2(\vecr,\vecr') = \frac{\delta^2S}{\delta\phi(\vecr)\delta\phi(\vecr')}\Big|_{\phi = i \psi}.
\end{equation}

Assuming that the ion-surface non-electrostatic potential (Fig. \ref{fig1}) is
shorter ranged than any other interaction, we can take the $a\to 0$ limit in the  continuum theory.
Then, the field action $S$ can be decomposed into separated volume (V) and surface (S) terms:
\begin{eqnarray}
\label{e3an}
\nonumber & &S\left[\phi(\vecr)\right] = \int_{V}\D\vecr\,\left( \frac{\beta\varepsilon(\vecr)}{8\pi}[\nabla \phi(\vecr)]^2
- 2\lambda\cos\left[\beta e\phi(\vecr)\right] \right)  \\
& &- \int_{S}\D^2r \, \lambda_s \left(\chi_+ \e^{-i\beta e\phi(z=0)} + \chi_-\e^{i\beta e\phi(z=0)} \right)  \, ,
\end{eqnarray}
where we introduced a phenomenological surface interaction strength, $\chi_{\pm}$, in order to describe the
specific short-range interaction between
ions and the surface
The $\chi_\pm$ parameter is explicitly connected with another surface interaction parameter, $\alpha_\pm$, by,
\begin{eqnarray}
\label{e3aa}
\chi_{\pm} \equiv a\left( \e^{-\beta\alpha_{\pm}} - 1 \right) \, ,
\end{eqnarray}
where $\alpha_{\pm}$, also known as {\it adhesivity}, is related to the average of the microscopic surface potential,
\begin{eqnarray}
\label{m10aaa}
\e^{-\beta\alpha_{\pm}} = a^{-1}\int_0^a \D z \, \e^{-\beta V_\pm(z)} \, .
\end{eqnarray}
We note that the above decomposition into bulk and surface terms enforces the partitioning of ions into
bulk and surface-residing. One thus needs to introduce also a specific surface fugacity, $\lambda_s$, that is different from the bulk one,
$\lambda_s = \lambda\exp[\varepsilon_w \lb (u_s-u_b)/2]$. This surface fugacity includes the ion self-energy at the surface, $u_s \neq u_b$,
as is elaborated in Sec.~II.B of Ref.~\cite{ionic_profiles}.

The ion surface properties as introduced above are
completely codified by the parameter $\chi_{\pm}$, Eq.~(\ref{e3aa}).
In the case of either repulsive or small attractive ion-surface interactions, $\chi_{\pm}$ is small,
and only terms of order ${\cal O}(\chi_{\pm})$ need to be considered.
This limit consistently leads to an effective Debye-H\"uckel (DH) theory as was elaborated in great detail in Refs.~\cite{JCP,EPL}.
However, for strong ion-surface interactions,
$\chi_{\pm}$ can be finite and one should generally keep all orders of $\chi_{\pm}$.
This further implies that the electrostatic potential cannot be linearized.
Rather, one needs to employ the full non-linear PB theory.

The one-loop grand-potential, Eq.~(\ref{e17}), is the starting point for our calculation. It constitutes of a mean-field  term and a fluctuation one.
The mean-field term, $\Omega_{\rm MF}$, is derived by substituting the field action, Eq.~(\ref{e3an}), into Eq.~(\ref{e17}),
\begin{eqnarray}
\label{e39}
\Omega_{_{\rm MF}} &=& \kbt S[\psi] = -\int\D\vecr\frac{\varepsilon(\vecr)}{8\pi}\left[\nabla\psi\right]^2
\nonumber\\
&-& 2n_b\kbt\int\D\vecr\cosh(\beta e\psi) \nonumber\\
&-& n_b\kbt \int\D^2r\,  \left[\chi_{+} \e^{-\beta e\psi_s} + \chi_{-} \e^{\beta e\psi_s)} \right]  \, ,
\end{eqnarray}
with the surface potential $\psi_s \equiv \psi(z=0)$.
The MF solution for $\psi$ is obtained from the saddle-point of the bulk part of the field action.
It leads to the standard PB equation, as is shown next. The fluctuation term, $\Omega_{_{\rm 1L}}$,
can be calculated by several routes \cite{Dean2}.
One method is based on the use of the {\it argument principle},
while a second one is based on the generalized Pauli -- van Vleck approach that calculates
the functional integral of a general harmonic
kernel. We shall proceed by employing the former methodology~\cite{JCP}.

\subsection{Mean Field}

The MF equation is derived from the saddle-point of the bulk field action.
In planar geometry, (Fig.~\ref{fig1}), this leads to the standard PB equation for $\psi(z)$
\begin{eqnarray}
\label{ep8}
\nonumber & &\psi^{\prime\prime}(z) = 0 \qquad\qquad\qquad\qquad\quad\,\,\,\, z<0 \, ,\\
& &\psi^{\prime\prime}(z) = \frac{8\pi e n_b}{\varepsilon_w} \sinh\left(\beta e\psi\right) \qquad\,\,\, z > 0\, ,
\end{eqnarray}
where $\psi'=\D\psi/\D z$, and we have used the translation symmetry in the transverse $(x,y)$ plane.
We also utilized the fact that in the MF approximation the fugacities are equal to the bulk salt concentration~\cite{ionic_profiles,JCP}.

The surface part of the saddle-point then gives a non-conventional boundary condition:
\begin{eqnarray}
\label{e31}
\nonumber \varepsilon_w\psi^\prime|_{_{0^+}} - \varepsilon_a \psi^\prime|_{_{0^-}}
= -4\pi e n_b \left( \chi_{+}\e^{-\beta e\psi_s} - \chi_{-}\e^{\beta e\psi_s} \right)  \, , \\
\end{eqnarray}
where $\psi_s$ is the surface potential and $\psi^\prime|_{_{0^\pm}}$ are its left and right first derivatives at $z\to 0$.
From the above equation we can define an {\it effective} surface charge density,
$\sigma_{\rm eff}$,  induced by the surface potential $\psi_s$,
\begin{equation}
\label{e32}
\sigma_{\rm eff}(\psi_s)= e n_b \left(\chi_{+}\e^{-\beta e\psi_s} - \chi_{-}\e^{\beta e\psi_s} \right) \, .
\end{equation}

Using the fact that $\psi$ vanishes at $z\rightarrow\pm\infty$, we obtain the usual relation~\cite{Safynia}:
\begin{eqnarray}
\label{e32a}
\nonumber\beta e\psi_s &=& 2 \ln\left( \frac{1+\eta}{1-\eta} \right)  \qquad\qquad\,\,\, z<0 \, , \\
\beta e\psi(z) &=& 2 \ln\left( \frac{1+\eta\e^{-\kd z}}{1-\eta\e^{-\kd z}} \right)  \qquad z\geq0 \, .
\end{eqnarray}
The parameter $0\le \eta\le 1$ is found by substituting $\psi$
from the above equation into the boundary condition at $z = 0$, Eq.~(\ref{e31}).
In addition, we have introduced the standard inverse Debye length, $\kd = \ld^{-1}=\sqrt{8\pi \lb n_b}$,
and assumed that  $\left|\chi_{+}\right| > \left|\chi_{-}\right|$, implying a positive effective surface charge and a positive surface potential.
For the opposite case of $\left|\chi_{+}\right|  < \left|\chi_{-}\right|$, one has to make the substitution $\eta \to -\eta$ in Eq.~(\ref{e32a}).

Inserting the solution of Eq.~(\ref{e32a}) into the boundary condition, Eq.~(\ref{e31}), yields an equation for $\eta$:
\begin{eqnarray}
\label{e32b1}
\nonumber \eta^4 &+& \eta^3\left( 2\kd\lgc - 4\Delta\chi \right) + 6\eta^2 \\
& & \quad - \, \eta\left( 2\kd\lgc + 4\Delta\chi \right) + 1 = 0 \, ,
\end{eqnarray}
where
\begin{eqnarray}
\label{e32b11}
\Delta\chi \equiv \left|\frac{\chi_{+} + \chi_{-} }{ \chi_{+} - \chi_{-} }\right|  \quad ; \quad
\lgc \equiv \frac{1}{2\pi\lb n_b\left|\chi_{+} - \chi_{-} \right|} \, .
\end{eqnarray}
Here $\Delta\chi$ is a modified (dimensionless) surface interaction strength, Eq.~(\ref{e3aa}), and $\lgc$ plays a similar role as the usual
Gouy-Chapman length~\cite{andelman2005}.
Note that the above equation applies equally to the case $\left|\chi_{+}\right|  < \left|\chi_{-}\right|$.

Keeping only linear terms in $\chi_{\pm}$ then leads to the regular Debye-H\"uckel (DH) solution.
For small enough bias, $\left|\chi_{+} - \chi_{-} \right| \to 0$, we have $\kd\lgc,\Delta\chi \gg 1$ yielding $\eta\ll1$,
and one can approximate the PB equation to order ${\cal O}\left(\eta\right)$ as~\cite{EPL}:
\begin{eqnarray}
\label{e32c}
\nonumber& &\beta e\psi_s = \frac{2}{2\Delta\chi+\kd\lgc}  \qquad\quad z<0 \, , \\
& &\psi(z) = \psi_s\e^{-\kd z}  \qquad\qquad\qquad z\geq0 \, .
\end{eqnarray}
%
If one furthermore assumes $\Delta\chi \ll \kd\lgc$, which corresponds to linearization in $\chi_{\pm}$,
the  DH solution is recovered~\cite{JCP}
\begin{eqnarray}
\label{e32d}
\beta e\psi_s = -\frac{2}{\kd\lgc} \, .
\end{eqnarray}
%

When $\chi_{-} + \chi_{+} > 0$, but either $\chi_{-} < 0$ or $\chi_{+} < 0$, the electrostatic potential
might be large and further considerations are required.
We assume, without loss of generality, $\left|\chi_+\right| > \left|\chi_-\right|$,
such that the effective surface charge is positive and $\chi_{-} < 0$.
Because $\chi_{-} < 0$, one should only keep terms to order ${\cal O}(\chi_{-})$.

In Appendix~B, we give further details on the complex expansion to first-order in $\chi_{-}$ that is used for our fitting procedure (see Section~IV).
However, in this subsection we only show the compact results obtained for $\chi_{-}=0$ (zeroth-order in $\chi_{-}$),
which is a good approximation when $|\chi_{+}|\gg|\chi_{-}|$.
Taking the zeroth order in $\chi_{-}$ yields $\Delta\chi \to 1$, $\lgc \to 1 / (2\pi\lb n_b \chi_{+})$,
and Eq.~(\ref{e32b1}) for $\eta$ takes a simpler form,
\begin{eqnarray}
\label{e32b21}
\nonumber\eta^3 + \eta^2\left( 2\kd\lgc - 3 \right) +\, \eta\left( 2\kd\lgc + 3 \right) - 1 = 0 \, . \\
\end{eqnarray}
The electrostatic potential, $\psi(z)$, is then derived by substituting $\eta$ of Eq.~(\ref{e32b21}) into Eq.~(\ref{e32a}).
Hereafter, we focus on the case with $\chi_{\pm} > 0$, which is equivalent to $\alpha_\pm<0$, meaning that both ions are attracted to the surface.

\subsection{One-Loop Correction}

In this section we follow the one-loop calculation described in Ref.~\cite{JCP} and will not dwell much on its details.
As discussed above, the one-loop correction to the grand-partition function, $\Omega_{_{\rm 1L}}$, can be rewritten with the help of the argument principle ~\cite{podgornik1989,attard,Dean2}, converting the discrete sum of the eigenvalues of the Hessian into
the logarithm of the {\it secular determinant} $D(k)$:
\begin{eqnarray}
\label{e42}
\Omega_{_{\rm 1L}} &=& \half\kbt\, {\rm Tr}\ln\left(H_2(\vecr,\vecr') \right) = \nonumber\\
& =& \frac{A\kbt}{8\pi^2} \int\D^2k\,\ln\left(\frac{D(k)}{D_{\rm free}(k)}\right) \, ,
\end{eqnarray}
where the integrand depends on the ratio $D(k)/D_{\rm free}(k)$,
and $D_{\rm free}$ is the reference secular determinant for a `free' system without ions.
The secular determinant is defined as~\cite{fdet}
\begin{eqnarray}
\label{e47}
D = \det\left[ M + N\Gamma(L)\Gamma^{-1}(0) \right] \, ,
\end{eqnarray}
with the matrix $\Gamma(z)$:
\begin{eqnarray}
\label{e48}
\Gamma(z) = \begin{pmatrix}
        h &  g\\
        \partial_z h &  \partial_z g
    \end{pmatrix} \, .
\end{eqnarray}

The two functions, $h(z)$ and $g(z)$, are the two independent solutions of the Hessian eigenvalue equation for zero eigenvalue,
\begin{equation}
\label{e43}
\left(\frac{\partial^2}{\partial z^2} -  k^2 -  \kd^2\cosh\left[\beta e\psi(z)\right]\right)u(z)=0 \, .
\end{equation}
The corresponding boundary condition of Eq.~(\ref{e43}) at $z=0$ is:
\begin{eqnarray}
\label{e44}
& & \varepsilon_w \partial_z{u}(z=0^+) - \varepsilon_a \partial_z{u}(z=0^-) = \omega {u}(0)\, ,
\end{eqnarray}
where we define,
\begin{eqnarray}
\label{e44a}
\omega \equiv  \half\varepsilon_w\kd^2 \left( \chi_{+} \e^{-\beta e\psi_s} + \chi_{-} \e^{\beta e\psi_s} \right) \, .
\end{eqnarray}
The two matrices $M$ and $N$ are obtained from writing the boundary condition in a matrix form (see Ref.~\cite{JCP}), yielding
\begin{eqnarray}
\label{s5}
M = \begin{pmatrix}
        -\omega - \varepsilon_a k &  \varepsilon_w \\
        0 & 0
    \end{pmatrix} \,\, ; \,\,
N = \begin{pmatrix}
        0 & 0 \\
        0 & 1
    \end{pmatrix} \, .
\end{eqnarray}

Using the expression of the MF potential, $\cosh(\beta e \psi_s)=2\coth^2[\kd(z+\zeta)]-1$
with $\zeta \equiv -(\ln\eta) / \kd$,
the two independent solutions of Eq.~(\ref{e43}) can be written as~\cite{lau2008}:
\begin{eqnarray}
\label{e46}
h(z) &=& \e^{p z} \left( 1 - \frac{\kd\coth\left[\kd(z+\zeta)\right]}{p} \right)\nonumber \, , \\
g(z) &=& \e^{-p z} \left( 1 + \frac{\kd\coth\left[\kd(z+\zeta)\right]}{p} \right) \, ,
\end{eqnarray}
where $p^2=k^2+\kd^2$.
By substituting Eq.~(\ref{e46}) into Eq.~(\ref{e47}), it is straightforward to compute the secular
determinant in the thermodynamical limit, $L\to\infty$. Using the limiting behaviors $g(L)\simeq g'(L)
\simeq 0$, $h(L)\simeq \exp(pL)(1-\kd/p)$ and $h'(L)\simeq ph(L)$, we obtain
\begin{eqnarray}
\label{e53a}
\nonumber& D(k)\simeq &-\frac{ph(L)}{2k^2}\left[ pg(0)\left( \omega + \varepsilon_a k + \varepsilon_w p \right)
\right. \\
& &\left.+\, \varepsilon_w \kd^2( \coth^{2}(\kd \zeta)-1) \right] \, .
\end{eqnarray}

In the DH regime, $\eta\ll 1$ and $\zeta \gg 1$. Hence, $D(k)$ reduces to
\begin{eqnarray}
\label{e53}
D(k) \simeq -\half\left[\omega + \varepsilon_a k + \varepsilon_w p \right]\e^{pL} \, ,
\end{eqnarray}
and $\omega$ reduces to $\half\varepsilon_w\kd^2 \left( \chi_{+} + \chi_{-}  \right)$.
This is exactly the DH result, which has been already obtained in Ref.~\cite{EPL}.

\begin{figure}[th]
\center
\includegraphics[scale=0.7,draft=false]{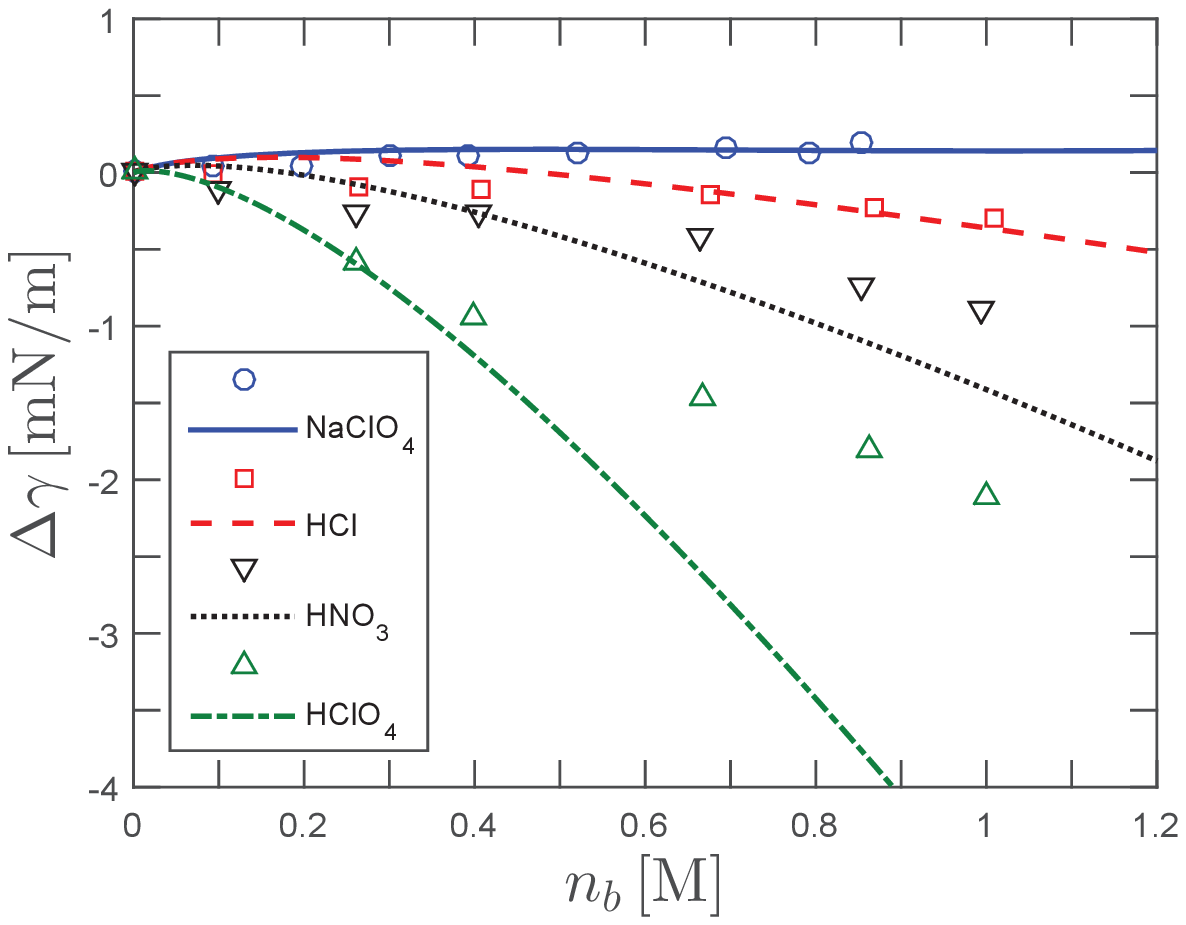}
\caption{\textsf{(color online).
Comparison of the fitted surface tension, $\Delta\gamma$, with experiments as a function of salt concentration, $n_b$,
at the air/water interface.  Experimental data are taken from
Ref.~\cite{Pugh} for the acids:  ${\rm HCl}$, ${\rm HNO_3}$, and ${\rm HClO_4}$,
and from Refs.~\cite{OxyExp,levin2010} for ${\rm NaClO_4}$.
The adhesivity values of $\alpha_{_{\rm H}}$ and $\alpha_{_{\rm ClO_4}}$
are found by first fitting ${\rm HCl}$ and ${\rm NaClO_4}$, while taking $\alpha_{_{\rm Cl}} = 0.09 \, \kbt$
and $\alpha_{\rm Na} = 0.11 \, \kbt$~\cite{JCP}.
We then use the values of $\alpha_{_{\rm H}}$ and $\alpha_{_{\rm ClO_4}}$ and the previously obtained
$\alpha_{_{\rm NO_3}} = -0.05 \, \kbt$\,\cite{JCP} to plot our predictions for the surface tension of ${\rm HClO_4}$  and ${\rm HNO_3}$.
The fitted adhesivity values are shown in Table~I. Other parameters are $T=300$\,K, $\varepsilon_w=80$ (water) and
$\varepsilon_a=1$ (air). }}
\label{fig2}
\end{figure}

\section{Surface Tension}

We can apply the formalism that was derived in the previous section to calculate the excess surface tension, $\Delta\gamma=\gamma-\gamma_{_{\rm A/W}}$,
which is the excess ionic contribution to the surface tension with respect to the surface tension between pure water and air, $\gamma_{_{\rm A/W}}$.
The surface tension can be calculated by using the Gibbs adsorption isotherm or, equivalently, by taking the difference between the
Helmholtz free-energy of an air/water system of longitudinal extent $2L$ (see Fig.~\ref{fig1})
and the sum of the Helmholtz free energies of the two corresponding bulk phases (each of longitudinal extent $L$):
\begin{eqnarray}
\label{st1}
\Delta\gamma&= &\left[F(2L) - {F}^{\rm (air)}(L)- { F}^{\rm (B)}(L)\right]/A \, .
\end{eqnarray}
The three Helmholtz free energies, $F(2L), {F}^{\rm (air)}(L)$, and ${ F}^{\rm (B)}(L)$,
have yet to be calculated explicitly.

The definition of the Helmholtz free energy is
\begin{eqnarray}
\label{e55a}
F = \Omega + \mu N + \mu_s N_s  \, ,
\end{eqnarray}
where the number of ions on the surface, $N_{s} = -\lambda_{s} \partial\Omega/\partial\lambda_{s}$.
Because $F$ is independent on the fugacities~\cite{JCP,demery},
the MF value (zeroth-loop order) of the fugacities, $\lambda = \lambda_s = n_b$, can be used.

\begin{table} [h]
\centering
\begin{tabular}{                  c                                       }
\hline
\hline
\qquad\qquad\qquad\qquad\quad\,\,\,\,\,\,  air/water   \,\,\,\,\,\,\quad\qquad\qquad\qquad\qquad  \\
\end{tabular} \\
\begin{tabular}{ c  c  c  c  c  c }
 \hline
  &  $a$    &   $\chi_{-}$  &    $\chi_{+}$  &   $\alpha_{-}$  &  $\alpha_{+}$  \\

  HCl              &  4.32 & -0.35 & 4.34  & 0.09  & -0.70  \\
  ${\rm HNO_3}$    &  4.35 & 0.21  & 4.37  & -0.05 & -0.70  \\
  ${\rm HClO_4}$   &  4.38 & 2.43  & 4.40  & -0.44 & -0.70   \\
  ${\rm NaClO_4}$  &  6.96 & 3.86  & -0.73 & -0.44 & 0.11 \\

\end{tabular} \\
\begin{tabular}{                  c                                       }
\hline
\hline
\qquad\qquad\qquad\qquad\quad\,\,\,\,\,  oil/water   \,\,\,\,\,\,\quad\qquad\qquad\qquad\qquad  \\
\end{tabular} \\
\begin{tabular}{ c  c  c  c  c  c }
  \hline
  &  $a$    &   $\chi_{-} $  &    $\chi_{+}$  &   $\alpha_{-}$  &  $\alpha_{+}$  \\

  KCl              &  6.63 & 0.48  & -0.91 & -0.07 & 0.15  \\
  KBr              &  6.61 & 1.64  & -0.91 & -0.22 & 0.15  \\
  KI               &  6.62 & 5.40  & -0.91 & -0.60 & 0.15  \\

  \hline \hline
\end{tabular}
\caption{\textsf{
Fitted values of the phenomenological surface interaction strength, $\chi_{\pm}$ (in \AA), and the corresponding
microscopic adhesivity, $\alpha_{\pm}$ (in $\kbt$),
at the air/water and dodecane/water interfaces. The $\alpha_{\pm}$ are obtained by the procedure elaborated in the text.
It includes predictions for ${\rm HClO_4}$ and ${\rm HNO_3}$. The radii, $a$ (in \AA), for all ions (except ${\rm H^+}$) are taken
from Ref.~\cite{IonRadii}. The effective ${\rm H^+}$ radius is taken from Ref.~\cite{Marcus}.
Note that all numerical values in the table and throughout the paper are rounded to two decimal places.}}
\label{table2}
\end{table}

For convenience, we separate the volume and surface contributions of the Helmholtz
free energy, $F=F_V+F_A$. The volume part, $F_V$, is written to the one-loop order~\cite{JCP} using Eqs.~(\ref{m3a}) and (\ref{e55a}):
\begin{eqnarray}
\label{e55}
\nonumber\frac{F_V}{V} &\simeq &\frac{\Omega_{_{\rm MF}}}{V} + 2\kbt n_b\ln (n_b a^3) - \frac{\kbt}{12\pi}\kd^3 \\
& & +\frac{\kbt}{8\pi}\kd^2\Lambda - \half e^2n_b u_{b}  \, .
\end{eqnarray}
Here we introduced the UV cutoff $\Lambda = 2\sqrt{\pi}/a$, where $a$ is the average minimal distance of approach between ions.
This cutoff is commonly used to avoid spurious divergencies arising when ions are assumed to be point-like (for further details see Ref.~\cite{ionic_profiles}).
In addition, we take the $\Lambda\rightarrow\infty$ limit and neglect all terms of order ${\cal O}(\Lambda^{-1})$.

The first two terms in $F_V$ are the MF grand potential, Eq.~(\ref{e39}), and the usual MF entropy contribution.
The third term is the well-known DH volume fluctuation term~\cite{Debye1923},
while the fourth and fifth terms are the bulk self-energies of the ions (diverging with the UV cutoff), which cancel each other exactly.

The surface part, $F_A$, is calculated solely from the one-loop correction:
\begin{eqnarray}
\label{e55aa}
\frac{F_A}{A} &=& \frac{\kbt}{4\pi}\int_0^{\Lambda}\D k\, k \Bigg[\ln \left(\frac{p-\kd}{k^3(\varepsilon_w + \varepsilon_a)}\right) \\
&+& \ln \Bigg(\big[p+\kd\coth(\kd \zeta)\big] \big[\omega +\varepsilon_a k +\varepsilon_w p\big] \nonumber\\
\nonumber&  &~~~~~~~ + \varepsilon_w\kd^2 (\coth^2(\kd \zeta)-1)\Bigg)\Bigg] - \half \, e^2 N_s u_{s}/A \, ,
\end{eqnarray}
where the last term in the above equation is proportional to the ion self-energy on the surface, $u_s$, which diverges with the cutoff.
This last term cancels with the leading divergence of the integral at the  $\Lambda\to \infty$ limit (just like the bulk one).

The bulk electrolyte free energy, $F^{\rm (B)}$, needed for Eq.~(\ref{st1}), is obtained from Eqs.~(\ref{e55}) and (\ref{e55aa})
in the same way as described in Sec.~IV of Ref.~\cite{JCP}.
In addition, the Helmholtz free energy of the air phase is equal to zero, $F^{\rm (air)}(L) = 0$, because there are no ions in the air phase.

\subsection{Mean Field}

Using the results for the three free-energies, we calculate the surface tension to one-loop order,
$\Delta\gamma \simeq \Delta\gamma_{_{\rm MF}} + \Delta\gamma_{_{\rm 1L}}$.
The mean-field (MF) part of the surface tension is derived using  $\psi(z)$
of Eqs.~(\ref{e32a}) and (\ref{e32b1}),
%
%
\begin{eqnarray}
\label{e60}
& & \Delta\gamma_{_{\rm MF}} = -\kbt n_b\left( \chi_{+}\e^{-\beta e \psi_s} + \chi_{-}\e^{\beta e \psi_s} \right) \\ \nonumber \\
\nonumber& &+ \int_{-\infty}^{\infty}\!\!\!\!\!\D z \left[-\frac{\varepsilon_w}{8\pi}\left(\frac{d\psi}{dz}\right)^2 + 2\kbt n_b\left(1 - \cosh\beta e\psi\right)\right] \, .
\end{eqnarray}
In the aqueous phase $z>0$, the first integration of Eq.~(\ref{e32a}) gives
\begin{eqnarray}
\label{e61}
\beta e\psi' = -2\kd\sinh\left(\beta e\psi/2\right) \, ,
\end{eqnarray}
while for $z<0$ (air), $\psi' = 0$. By inserting $\psi'(z)$ into Eq.~(\ref{e60}) and integrating, we obtain the MF surface tension
\begin{eqnarray}
\label{e62}
\nonumber \Delta\gamma_{_{\rm MF}} &=& -\kbt n_b \Bigg[ \chi_{+}\e^{-\beta e \psi_s} + \chi_{-}\e^{\beta e \psi_s}  \\
& & \quad+ 8 \kd^{-1} \Big(\cosh\left[\beta e\psi_s/2\right] - 1\Big)\Bigg] \,.
\end{eqnarray}
This expression is similar Eq.~(3.16) of Ref.~\cite{diamant1996}, where the surface tension was calculated for charged surfactants adsorbing
onto the air/water interface. It is worth noting that by taking $\chi_{\pm} \to 0$, the surface potential $\psi_s$
vanishes and consequently the entire MF contribution to the surface tension is zero.
This leads back to the OS result which is a fluctuation term.

\subsection{One-loop Correction}

The one-loop correction to the surface tension takes the following form:
\begin{eqnarray}
\label{e64a}
\nonumber\Delta\gamma_{_{\rm 1L}} &=& \frac{\kbt}{8\pi}\int_0^{\Lambda}\D k\, k\ln \Bigg[ \frac{(p - \kd)}{k^3\left(\varepsilon_w + \varepsilon_a \right)^2} \\
\nonumber&\times& \Big( \varepsilon_w\kd^2 \sinh^{-2}(\kd \zeta)  \\
\nonumber&+& \left[p+\kd\coth(\kd \zeta)\right]\left[\omega + \varepsilon_a k + \varepsilon_w p\right] \Big)^2 \\
\nonumber&\times& \left(p^2 + p\kd \coth(\kd \zeta) + \half\kd^2\sinh^{-2}(\kd \zeta) \right)^{-1} \Bigg] \\
&-&\frac{\kbt}{4\pi}\frac{\omega\Lambda}{\varepsilon_w+\varepsilon_a} \, .
\end{eqnarray}
%

Taking the limit of $\eta \ll 1$ (or $\zeta= -(\ln\eta) / \kd \gg 1$) gives the linearized fluctuation contribution as obtained in Refs.~\cite{EPL,JCP}:
\begin{eqnarray}
\label{so3}
\nonumber \frac{8\pi}{\kbt}\Delta\gamma_{1} &\simeq& -
\left(\frac{\varepsilon_{w}-\varepsilon_a}
{\varepsilon_w+\varepsilon_a}\right)\frac{\kappa_D^2}{2} \left[ \ln\left(\frac{1}{2}\kd\lb\right)
- \ln\left(\frac{1}{2}\lb\Lambda \right)\right. \nonumber\\
&-&  \left. \frac{2\omega^2}{\kappa_D^2(\varepsilon_w^2-
\varepsilon_a^2)} \ln\left(\kappa_D\Lambda^{-1}\right) \right] \, ,
\end{eqnarray}
where only $\Lambda$-dependent terms are shown.
The first term in Eq.~(\ref{so3}) is the well-known OS result~\cite{onsager_samaras,podgornik1988,dean2004}
and it varies as $\sim\kappa_D^2\ln(\kappa_D\lb)$. The second term is a correction due to the ion minimal distance
of approach, with the UV cutoff $\Lambda=2\sqrt{\pi}/a$, while the third term is a correction related to the adhesivity parameters through
$\omega(\alpha_\pm)$, Eq.~(\ref{e44a}).
For $\beta\alpha_{\pm} \ll 1$, the third term is negligible and, as expected, the derived surface tension agrees well with
the OS result.

\begin{figure*}[th]
\center
\includegraphics[scale=0.9,draft=false]{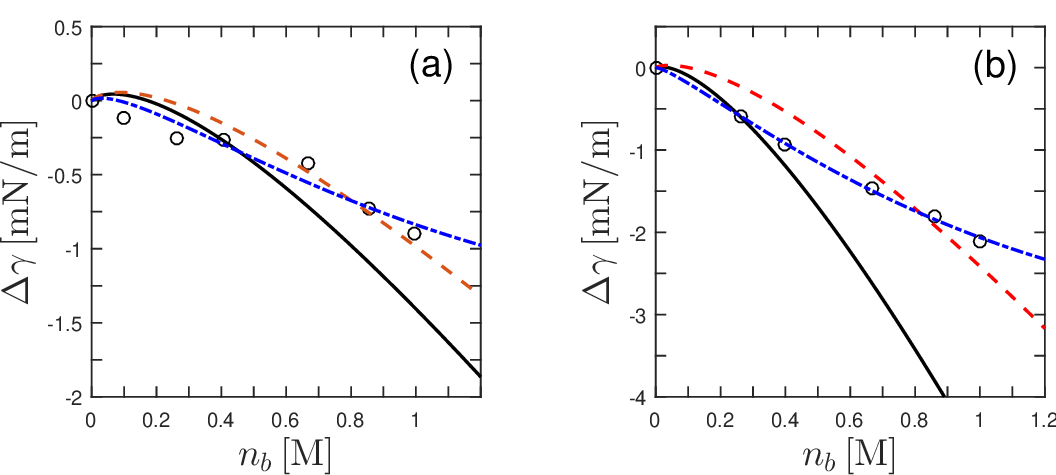}
\caption{\textsf{(color online).
Comparison of the calculated surface tension (black circles) with experiments at the air/water interface as function of ionic concentration, $n_b$,
for ${\rm HNO_3}$ (a) and ${\rm HClO_4}$ (b).
The predicted black solid line is calculated from the procedure elaborated in the text for
$\alpha_{_{\rm ClO_4}} = -0.44 \, \kbt$ and $\alpha_{_{\rm NO_3}} = -0.05 \, \kbt$ (see Table~\ref{table2}).
The red dashed line is a one-parameter fit for $\alpha_{_{\rm ClO_4}}$ and $\alpha_{_{\rm NO_3}}$, yielding less negative or even positive adhesivity values:
$\alpha_{_{\rm ClO_4}} = -0.17 \, \kbt$ and $\alpha_{_{\rm NO_3}} = 0.01 \, \kbt$.
For both curves, we use $\alpha_{_{\rm H}} = -0.70 \, \kbt$ (see Table~I).
The third, blue dash-dotted line, is the ``best fit" (2-parameter fit) yielding:
$\alpha_{_{\rm H}} = -1.11 \, \kbt$ and $\alpha_{_{\rm NO_3}} = 0.17 \, \kbt$ for ${\rm HNO_3}$,
and $\alpha_{_{\rm H}} = -1.57 \, \kbt$ and $\alpha_{_{\rm ClO_4}} = 0.17 \, \kbt$ for ${\rm HClO_4}$.
Other parameters are as in Fig.~\ref{fig2}.
}}
\label{fig3}
\end{figure*}

\section{Comparison with Experiments}

We compare the numerical results for the surface tension (computed from the one-loop fluctuation correction of the MF results),
$\Delta\gamma = \Delta\gamma_{_{\rm MF}}+ \Delta\gamma_{_{\rm 1L}}$, with experimental data.
For the case where $\chi_{\pm}>0$, we use Eq.~(\ref{e62}) for $\Delta\gamma_{_{\rm MF}}$ and Eq.~(\ref{e64a})
for $\Delta\gamma_{_{\rm 1L}}$. On the other hand, if either $\chi_+$ or $\chi_-$ is negative,
we expand to first order in the negative $\chi$, as shown in Appendix~B and is explained in the paragraph after Eq.~(\ref{m10aaa}).
Then, the MF term, $\Delta\gamma_{_{\rm MF}}$, is derived from Eq.~(\ref{e62a})
and $\Delta\gamma_{_{\rm 1L}}$ is obtained from Eqs.~(\ref{e64a1})-(\ref{e64a1n}). For simplicity, we take the range of the ion-specific
surface potential to be equal to $a$, the average minimal distance between cations and anions in water, yielding
$a = r_{+}^{\rm hyd} + r_{-}^{\rm hyd}$, with the hydrated radii taken from literature~\cite{IonRadii,Marcus}.

Our fitting procedure is centered on obtaining the best fitted values for the phenomenological adhesivities, $\alpha_{\pm}$.
These adhesivities are extracted from one of the fits and uniquely determine the adhesivity value of the specific ion/interface system for the other fits.
This procedure allows us to make predictions for other salt solutions.
Note that the surface tension is symmetric with respect to exchanging the role of cations and anions.
This means that the two-parameter fit with $\alpha_{\pm}$ will always give two equivalent results, $\alpha_{+}
\leftrightarrow\alpha_{-}$. An alternative fitting procedure was used in Ref.~\cite{JCP},
for a different case in which both adhesivities are small, $|\beta\alpha_{\pm}| \ll 1$. Then, $\alpha^{*} = \alpha_- + \alpha_+$ can be introduced as a single
fit parameter yielding almost equivalent results.

In Fig.~\ref{fig2}, we compare the analytical results for the surface tension of acids at the air/water interface
with experimental data.  The experimental data show that
the surface tension decreases or slightly increases with ionic concentration.
This indicates a relatively strong ion-surface interaction that cannot be treated within the DH linear theory,
and is consistent with our starting point.
The three HX acids~\cite{Pugh}, with ${\rm X}=$ ${\rm Cl^-}$, ${\rm NO_3^-}$, or ${\rm ClO_4^-}$,
and a salt with an oxy anion, ${\rm NaClO_4}$ ~\cite{OxyExp,levin2010} are used in the comparison. We fit their surface tension curves
with $\alpha_{_{\rm Na}}$, $\alpha_{_{\rm Cl}}$, and $\alpha_{_{\rm NO_3}}$, which were derived in our previous work~\cite{JCP}.

In the fitting procedure, we first fit the surface tension of  ${\rm HCl}$ and ${\rm NaClO_4}$ in order to find $\alpha_{_{\rm H}}$ and $\alpha_{_{\rm ClO_4}}$.
This allows us to predict the surface tension of ${\rm HNO_3}$ and ${\rm HClO_4}$.
The ionic radii for all ions except hydrogen are taken from Ref.~\cite{IonRadii},
and the hydrogen effective radius in water\footnote{The effective hydrogen radius includes its various complexations with water molecules.}
is taken from Ref.~\cite{Marcus}.
The surface tension for ${\rm NaClO_4}$ and ${\rm HCl}$ is in very good agreement with experiments for the entire concentration range
(up to $\sim$1\,M), while for ${\rm HNO_3}$ and ${\rm HClO_4}$ the surface tension shows deviation from experiments at high
concentrations ($\gtrsim0.7$\,M for ${\rm HNO_3}$ and $\gtrsim0.4$\,M for ${\rm HClO_4}$).

In Fig.~\ref{fig3} we plot three fitting curves for ${\rm HNO_3}$ in (a) and for ${\rm HClO_4}$ in (b).
The first plot is our prediction as seen in Fig.~\ref{fig2},
the second uses $\alpha_{_{\rm H}}$ and then fits the best value for $\alpha_{_{\rm NO_3}}$ and $\alpha_{_{\rm ClO_4}}$, while
the third is the ``best fit" optimized for both $\alpha$ values. In the first two fits, we use $\alpha_{_{\rm H}} = -0.70\, \kbt$ of Table~I.
The second curve fits rather well, certainly  better than the prediction of the first curve, and corresponds to less negative adhesivity
values: $\alpha_{_{\rm NO_3}} = 0.01 \, \kbt$ (as opposed to $\alpha_{_{\rm NO_3}} = -0.05 \, \kbt$) and
$\alpha_{_{\rm ClO_4}} = -0.17 \, \kbt$ (as opposed to $\alpha_{_{\rm ClO_4}} = -0.44 \, \kbt$). The difference in the estimated
adhesivities between the first two fits implies the existence of a mechanism that will tend to diminish
their values, effectively excluding the ions from the surface. A possible source of this exclusion can be associated with
steric ion-ion repulsion  at the surface\footnote{This exclusion depends on ionic size and precludes unbound densities
of the adsorbed ions in the limit $\beta\alpha_{\pm} \to -\infty$, setting an upper bound corresponding to the close-packing configuration,
and is similar to systems with charge-regulated boundary condition~\cite{CR1,Safynia}.}.

\begin{figure}[th]
\center
\includegraphics[scale=0.7,draft=false]{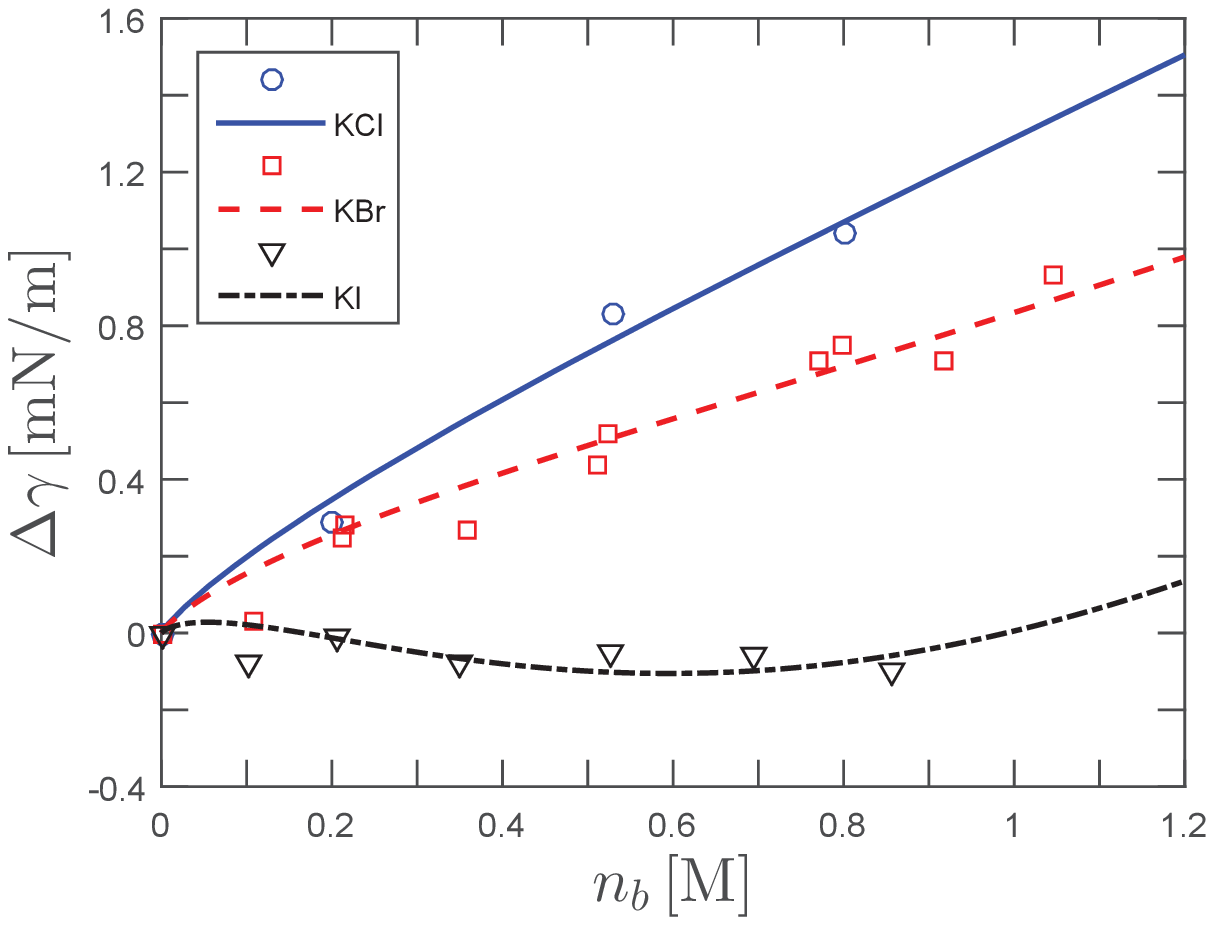}
\caption{\textsf{(color online).
Comparison of the calculated  surface tension with experimental data from Ref.~\cite{KExp}, as function of ionic concentration, $n_b$,
at the dodecane/water interface.
The three hilade/alkaline salts are ${\rm KCl}$, ${\rm KBr}$ and ${\rm KI}$.
The adhesivities values are extracted from first fitting the KI curve.
Then, we use the value of $\alpha_{_{\rm K}} = 0.15 \, \kbt$ and fit the surface tension of
the other two salts, ${\rm KBr}$ and ${\rm KCl}$.
The fitted adhesivity values, $\alpha_{\pm}$, are shown in Table~I.
Other parameters are as in Fig. 2, beside the dielectric constant of dodecane, $\varepsilon_a = 2$.
}}
\label{fig4}
\end{figure}

In addition, our approach  successfully applies to other types of liquid interfaces, such as oil/water.
This is demonstrated in  Fig.~\ref{fig4}, where we compare the calculated surface tension for dodecane/water interface with experiments.
The fits are done for three different salts having K$^+$ as their common cation, and they are in very good agreement with experiments.
The adhesivity values are obtained by first fitting the KI data. Then, this value of $\alpha_{_{\rm K}} = 0.15 \, \kbt$ is used
in order to fit the surface tension of the two homologous salts, KBr and KCl.
Notice that the adhesivity values for ${\rm KCl}$ and ${\rm KBr}$ are rather small
and, thus, are similar to the results of the linearized DH theory of Ref.~\cite{JCP}. However, $\alpha_{_{\rm I}} \simeq -0.6 \, \kbt$ is not
that small, and the corresponding fit for KI is greatly improved when compared to Ref.~\cite{JCP}.

Together with the previous results of Ref.~\cite{JCP}, we obtain an extended reverse Hofmeister series
with decreasing adhesivity strength at the air/water interface:
${\rm F}^{-} > {\rm IO_3}^{-}  > {\rm Cl}^{-} > {\rm BrO_3}^{-} > {\rm Br}^{-} > {\rm ClO_3}^{-} > {\rm NO_3}^{-} > {\rm I}^{-} > {\rm ClO_4}^{-}$,
while for cations the series is:
${\rm K}^{+} > {\rm Na}^{+} > {\rm H}^{+}$.
At the oil/water interface as in Fig.~\ref{fig4}, the same reversed Hofmeister series emerges with more attractive ion-surface interactions.
This effect is substantially stronger for the anions,
and might be connected with the stronger dispersion forces at the oil/water interface~\cite{ninham2001},
or change in the strength of hydrogen bonds close to the surface (see Ref.~\cite{JCP} for further discussion).

\section{Conclusions}

Our present work complements previous results obtained for surface tension of weakly
adhering electrolytes~\cite{JCP,EPL},
and extend them to strong acids, bases, and other ions that strongly adsorb to the interface. This study is accomplished by considering
the full non-linear PB theory for mean-field and one-loop fluctuation correction, which is valid for any strength
of the ion-surface interaction (the surface adhesivity, $\alpha$, in our model). In particular, we were able to obtain {\it analytically}
the surface tension up to the one-loop order. As was explained before, the fluctuation correction is
paramount to this endeavour as it generalizes the OS argument, which is itself fluctuational in nature~\cite{JCP,EPL}.

The analytical expressions derived for the surface tension is applicable for any adhesivity values,
and reduces to results we derived previously for small adhering asymmetry ($\alpha_{+} \simeq \alpha_{-}$).
Nevertheless, we expect that for the extreme case of strong adhesivities
and high salt concentration, other effects such as ion-ion steric interactions, will play a role.
Our results for the surface tension are in accord with the reverse Hofmeister series at the oil/water interface
and extend the series to acids.

It is possible to generalize our model to include the surface tension of weak acids.
Conceptually, the main change will be that the molarity of H$^+$ is a function of the bulk concentration,
$n_b$ and pK, $[{\rm H}^{+}]=f(n_b,{\rm pK})$.
As written in the introduction, this task is rather simple if one takes the pK value to be constant throughout the solution~\cite{weak_acids}.
However, the corresponding equations that take fully into account the {\it local} acid dissociation reaction are more complex,
though imminently solvable (see Ref.~\cite{bulk_CR}).
Such a relation will be needed in order to compute the surface tension as a function of the experimental controlled molarity of the acid solution, $n_b$.

Finally, we note that ion-surface interactions are the core of the ionic-specific Hofmeister series.
This statement is based on the generality of our model, its natural inclusion of the OS result,
and the very good fit to experimental data.
With the same simple idea, and by merely taking into account the ion-surface specific interactions,
we were able to recover the reverse Hofmeister series
and calculate the surface tension for weakly adsorbed ions at a surface~\cite{EPL} or within a proximal layer~\cite{JCP},
strongly adsorbed ions or acids (the present work), and ionic profiles in the vicinity of the interface~\cite{ionic_profiles}.
In the future, we hope that better understanding
of the behavior of ions at interfaces will rely on more refined models that will explore the microscopic
origin of the adhesivity parameter, $\alpha$.

\vskip 0.5cm

{\it Acknowledgements.~~~} We thank A. Cohen, R. M. Adar, H. Orland and H. Diamant,
for useful discussions and numerous suggestions.
This work was supported in part by the Israel Science Foundation (ISF) under Grant No. 438/12
and the US-Israel Binational Science Foundation (BSF) under Grant No. 2012/060, and the ISF-NSFC joint research program under Grant No. 885/15.
R.P. would like to acknowledge the hospitality of the Tel Aviv University during his multiple visits there.

\appendix

\section{Adding External Surface Charge}

Throughout this work we considered surfaces that are characterized by an adhesivity parameter, $\alpha$,
which is responsible for the ionic profiles at the surface/interface vicinity.
Here, we extend these results and include fixed charge groups of density $\sigma$ on the surface.
Including $\sigma$, together with the surface adhesivity $\alpha_+$, modifies
Eq.~(\ref{e3an}) into the form
\begin{eqnarray}
\label{e67}
\nonumber& S &= \int_{V}\D\vecr\,\left(\frac{\beta\varepsilon_w}{8\pi}[\nabla \phi(\vecr)]^2 - 2\lambda\cos\left[\beta e\phi(\vecr)\right] \right)  \\
& &- \int\D\vecr \left[\lambda_s\chi_+ \,\e^{-i\beta e\phi(\vecr)} - i\beta\sigma\phi(\vecr)\right]\delta(z)  \, .
\end{eqnarray}
For simplicity, we only consider the cation adhesivity ($\chi_-=0$), and assume positive adsorption for the cations, such that $\chi_+>0$.

The MF equation, Eq.~(\ref{ep8}), does not change, but the boundary condition at $z=0$ is modified:
\begin{eqnarray}
\label{e68}
\varepsilon_w\psi_2^\prime|_{_{0^+}} - \varepsilon_a \psi_1^\prime|_{_{0^-}} = -4\pi \left(\sigma + \sigma_0 \e^{-\beta e\psi_s}\right) \, ,
\end{eqnarray}
with $\sigma_0 = e n_b \chi_{+}$.
The MF solution, Eq.~(\ref{e32a}), depends on $\eta$, which by itself is derived from the boundary condition, Eq.~(\ref{e68}),
\begin{eqnarray}
\label{e69}
\nonumber \eta^3 + \eta^2\left( 2\kd\lsig -\Delta\sigma \right) + \eta\left(  2\kd\lsig + \Delta\sigma \right) - 1 = 0 \, . \\
\end{eqnarray}
In the above equation we define $\Delta\sigma \equiv (3\sigma_0 - \sigma)/(\sigma_0 + \sigma)$ and $\lsig \equiv e/(2\pi \lb|\sigma_0 + \sigma|)$, where the latter plays the role of the Gouy-Chapman length.
This is the solution for $\sigma_0 + \sigma > 0$, while
for $\sigma_0 + \sigma < 0$, one has to take $\eta \to -\eta$ and $\lsig \to -\lsig$.

By taking $\sigma=0$, we recover the case of no fixed surface charges, Eq.~(\ref{e32a}), for $\chi_- = 0$.
On the other hand, if we take $\chi_+=0$, one obtains the well-known equation for $\eta$ for a single charged surface in contact with an electrolyte~\cite{Safynia,andelman2005}:
\begin{eqnarray}
\label{e69b}
\eta^2 + 2\kd\lsig\,\eta - 1 = 0 \, .
\end{eqnarray}
When $\left|\sigma_{ 0} + \sigma \right| \ll 1$, it can be shown that $\eta \ll 1$.
Taking only terms of order ${\cal O}\left(\eta\right)$ yields:
\begin{eqnarray}
\label{c5}
\nonumber & &\beta e\psi_s \simeq \frac{2}{\kd\lsig + \Delta\sigma/2 }  \qquad\qquad \!\!\!\! z<0 \, , \\
& &\psi = \psi_s\e^{-\kd z} \qquad\qquad\qquad\qquad z\geq0 \, .
\end{eqnarray}
If both $\sigma$ and $\sigma_0$ are small, $\kd\lsig \gg \Delta\sigma$ and we recover the DH solution for an effective surface charge:
\begin{eqnarray}
\label{c5a}
\beta e\psi_s \simeq -\frac{2}{\kd\lsig} \, .
\end{eqnarray}

The free-energies of the bulk and air phases do not change, and the MF surface tension can be derived as before:
\begin{eqnarray}
\label{e71}
\nonumber\Delta\gamma_{_{\rm MF}} &=& -\kbt \Bigg[ n_b \chi_+ \,\e^{-\beta e \psi_s} - \beta\sigma\psi_s \\
& & \quad + \, 8 n_b  \kd^{-1} \Big(\cosh\left[\beta e\psi_s/2\right] - 1\Big)\Bigg].
\end{eqnarray}

The addition of fixed surface charge affects the one-loop correction only via the MF potential.
The one-loop surface tension, $\Delta\gamma_{_{1L}}$, can be derived from  Eq.~(\ref{e64a}),
by taking the MF potential obtained from  Eqs.~(\ref{e32a}) and (\ref{e69}).

It is clear that the addition of fixed surface charges only affect the MF surface tension, hence, it can be easily incorporated into our methodology.

\section{Strong Surface Potential with $\chi_-<0$}

In this appendix we compute the surface tension for the case in which either $\chi_+$ or $\chi_-$ is negative.
In such a case, the negative $\chi$ is always of the order of $a$.
Thus, in order to be consistent with the limit taken in Eq.~(\ref{e3an}), one must keep only linear terms of the negative $\chi$.

Without loss of generality we assume that $\left|\chi_+\right| > \left|\chi_-\right|$, such that the effective surface charge is positive.
In such a case, having a strong electric potential requires $\left|\chi_-/\chi_+\right|\ll 1$.
We write $\eta = \eta_0 + \left(\chi_-/\chi_+\right) \eta_1$, which implies that $\psi=\psi_0 + \left(\chi_-/\chi_+\right) \psi_1$,
and is consistent with the limit $a\to0$ of Eq.~(\ref{e3an}).
Using this expansion in Eq.~(\ref{e32a}) gives,
\begin{eqnarray}
\label{e32a1}
& & \beta e\psi_s \simeq \beta e\left(\psi_0^{(s)}+\frac{\chi_-}{\chi_+} \, \psi_1^{(s)}\right) \nonumber\\
\nonumber& & = 2 \ln\left( \frac{1+\eta_0}{1-\eta_0} \right) + \frac{2}{1-\eta_0^2} \,  \frac{\chi_-}{\chi_+} \, \eta_1 \qquad\qquad\qquad\qquad\!\!\! z<0 \, , \\
\nonumber& & \beta e\psi \simeq \beta e\left(\psi_0 + \frac{\chi_-}{\chi_+} \, \psi_1\right) \\
\nonumber& &= 2 \ln\left( \frac{1+\eta_0\e^{-\kd z}}{1-\eta_0\e^{-\kd z}} \right)
+ \frac{2\e^{-\kd z}}{1-\eta_0^2\e^{-2\kd z}} \,  \frac{\chi_-}{\chi_+} \, \eta_1  \qquad z\geq0 \, .\nonumber\\
\end{eqnarray}

Equation~(\ref{e32b1}) for $\eta$ takes a simpler form by using
$\Delta\chi \simeq 1 + 2\chi_-/\chi_+$ and $\lgc \simeq \lgc^{(0)} (1+\chi_-/\chi_+)$,
\begin{eqnarray}
\label{e32b21n}
\nonumber & &\eta_0^3 + \eta_0^2\left( 2\kd\lgc^{(0)} - 3 \right) +\, \eta_0\left( 2\kd\lgc^{(0)} + 3 \right) - 1 = 0 \, , \\
& &\eta_1 = \eta_0 \, \frac{4+\kd\lgc^{(0)}+\eta_0^2\left(4-\kd\lgc^{(0)}\right)}{2\left(\eta_0-1\right)^3+\kd\lgc^{(0)}\left(3\eta_0^2-1\right)} \, ,
\end{eqnarray}
%
where $\lgc^{(0)} \equiv 1 / (2\pi\lb \chi_{+}) $.

Substituting the MF potential of Eq.~(\ref{e32a1}), we write the MF surface tension, Eq.~(\ref{e62}), to first order in $\chi_- / \chi_+$ as
\begin{eqnarray}
\label{e62a}
\nonumber \Delta\gamma_{_{\rm MF}} &=& -\kbt n_b \Bigg[ \chi_{+}\e^{-\beta e \psi_0^{(s)}}
\!\!\!+ 8 \kd^{-1} \left( \cosh\left[\beta e\psi_0^{(s)}/2\right] - 1\right) \\
\nonumber&+& \frac{\chi_-}{\chi_+} \Bigg( \chi_+\left[ \e^{\beta e \psi_0^{(s)}} - \beta e\psi_1^{(s)} \e^{-\beta e \psi_0^{(s)}}  \right]  \\
& & \qquad + \, 4 \kd^{-1} \beta e\psi_1^{(s)}\sinh\left[\beta e\psi_0^{(s)}/2\right] \Bigg) \Bigg] \, .
\end{eqnarray}
%


In order to expand the one-loop surface tension, Eq.~(\ref{e64a}), to first order in $\chi_-/\chi_+$ we first write,
\begin{eqnarray}
\label{e64a1m}
\zeta &\simeq&  \zeta_0 + \frac{\chi_-}{\chi_+}  \zeta_1 = -\frac{\ln\eta_0}{\kd} - \frac{\chi_-}{\chi_+}\frac{\eta_1}{\kd\eta_0}
\end{eqnarray}
with
\begin{eqnarray}
\omega &\simeq& \omega_0 + \frac{\chi_-}{\chi_+} \omega_1 \\
\nonumber &=& \frac{\varepsilon_w}{2}\kd^2 \chi_{+} \e^{-\beta e\psi_0^{(s)}}
\left[ 1 + \frac{\chi_-}{\chi_+}\left( \e^{2\beta e\psi_0^{(s)}} - \beta e\psi_1^{(s)} \right) \right]\, .
\end{eqnarray}
Expanding  Eq.~(\ref{e64a}) to first order in $\chi_-/\chi_+$ and writing
$\Delta\gamma_{_{\rm 1L}} = \Delta\gamma_0^{_{\rm 1L}} + \left(\chi_-/\chi_+\right) \Delta\gamma_1^{_{\rm 1L}}$, we obtain:
\begin{widetext}
\begin{equation}
\begin{aligned}
\label{e64a1}
\Delta\gamma_0^{_{\rm 1L}} &= \frac{\kbt}{8\pi}\int_0^{\Lambda}\D k\, k\ln \Bigg[ \frac{(p - \kd)}{k^3\left(\varepsilon_w + \varepsilon_a \right)^2}
\times \Big( \varepsilon_w\kd^2 \sinh^{-2}(\kd \zeta_0) + P\omega_{kp} \Big)^2 \\
&\times \left(pP + \half\kd^2\sinh^{-2}(\kd \zeta_0) \right)^{-1} \Bigg] -\frac{\kbt}{4\pi}\frac{\omega_0\Lambda}{\varepsilon_w+\varepsilon_a}  \, ,
\end{aligned}
\end{equation}
\end{widetext}
where we defined for convenience two auxiliary variables
\begin{equation}
\begin{aligned}
& &\omega_{kp} = \omega_0 + \varepsilon_a k + \varepsilon_w p \, , \\
& &P = p+\kd\coth(\kd \zeta_0) \, ,
\end{aligned}
\end{equation}
and
\begin{widetext}
\begin{equation}
\begin{aligned}
\label{e64a1n}
\Delta\gamma_1^{_{\rm 1L}} &= \frac{\kbt}{8\pi}  \int_0^{\Lambda}\D k k \Bigg[
\omega_1 \Bigg(  4 p  \left[ \kd^2 + p^2 + 2 \kd p \coth\left(\kd\zeta_0\right) \right]
+ \frac{2 \kd^2 \left[ 2p+P \right]}{\sinh^2(\kd\zeta_0) }  \Bigg) \\
&-  \zeta_1 \frac{2 \kd^2}{\sinh^2(\kd\zeta_0)} \Bigg(  \varepsilon_w p^3 + k^2 \left( \varepsilon_a k + \omega_0 \right)\\ &+ 3\varepsilon_w \kd^2 p
+ 4\varepsilon_w\kd p^2 \coth(\kd\zeta_0) + \varepsilon_w\kd^2\frac{3p + \kd\coth(\kd\zeta_0)}{\sinh^2(\kd\zeta_0)}  \Bigg)  \Bigg] \\
&\times \Bigg(  2 p P^2\omega_{kp}
+ \kd^2 \sinh^{-2}(\kd\zeta_0) \left[  \omega_{kp} + 2 \varepsilon_w p   \right]P
+  \varepsilon_w \kd^4 \sinh^{-4}(\kd\zeta_0) \Bigg)^{-1}\\
&-~\frac{\kbt}{4\pi}\frac{\omega_1\Lambda}{\varepsilon_w+\varepsilon_a} \, .
\end{aligned}
\end{equation}
\end{widetext}
These analytical but rather complex expressions are used in the calculation of the surface tension throughout the paper
for the case in which either $\chi_+$ or $\chi_-$ is negative.



\end{document}